\documentclass[aps,prd,superscriptaddress,12pt,showpacs,notitlepage]{revtex4}
\usepackage{graphics}
\usepackage{graphicx}
\usepackage{amsfonts}
\usepackage{amsmath}

\newcommand{\be}{\begin{equation}}
\newcommand{\ee}{\end{equation}}
\newcommand{\bea}{\begin{eqnarray}}
\newcommand{\eea}{\end{eqnarray}}
\newcommand{\pa}{\partial}
\newcommand{\bb}{\bibitem}
 
\begin{document}
\title{BPS bounds in supersymmetric extensions of K field theories}
\author{C. Adam}
\affiliation{Departamento de F\'isica de Part\'iculas, Universidad de Santiago de Compostela and Instituto Galego de F\'isica de Altas Enerxias (IGFAE) E-15782 Santiago de Compostela, Spain}
\author{J.M. Queiruga}
\affiliation{Departamento de F\'isica de Part\'iculas, Universidad de Santiago de Compostela and Instituto Galego de F\'isica de Altas Enerxias (IGFAE) E-15782 Santiago de Compostela, Spain}
\author{J. Sanchez-Guillen}
\affiliation{Departamento de F\'isica de Part\'iculas, Universidad de Santiago de Compostela and Instituto Galego de F\'isica de Altas Enerxias (IGFAE) E-15782 Santiago de Compostela, Spain}
\author{A. Wereszczynski}
\affiliation{Institute of Physics,  Jagiellonian University,
       \\ Reymonta 4, 30 059 Krak\'{o}w, Poland}


\begin{abstract}
We demonstrate that in the supersymmetric extensions of a class of generalized (or K) field theories introduced recently, the static energy satisfies a BPS bound in each topological sector. Further, the corresponding soliton solutions saturate the bound. We also find strong indications that the BPS bound shows up in the SUSY algebra as a central extension, as is the case in the well-known supersymmetric field theories with standard kinetic terms.
\end{abstract}

\maketitle 

\section{Introduction}
If a quantum field theory is assumed to be applicable to physical processes at arbitrary energy scales, then both its field contents and possible terms contributing to the Lagrangian are quite constrained, mainly by the requirement of renormalizability. Recently, however, a different point of view has gained support, where the field theory under consideration is interpreted as a low-energy effective field theory which, at sufficiently high energies, is superseded by a more fundamental theory (string theory being the most prominent proposal). In this effective field-theory interpretation, the presence of non-renormalizable terms in the lagrangian just indicates the existence of a natural cutoff in the effective field theory, beyond which calculations within the effective field theory framework are no longer trustworthy, and effects of the fundamental theory have to be taken into account. The effective field theory point of view, therefore, allows to consider a much broader class of Lagrangians, which may, in a first instance, be rather general functions of the fields and their space-time derivatives. Allowing for higher than first derivatives in the Lagrangian, however, may introduce some further problems like, e.g., the necessity to introduce ghosts, so it is natural to consider a class of generalized field theories given by Lagrangians which depend in a Poincare-invariant way on the fields and on their first derivatives. Specifically, a broader class of kinetic terms, generalizing the standard quadratic kinetic terms, may be considered. These theories with generalized kinetic terms (termed K field theories) have been studied with increasing effort in the last years, especially in the context of cosmology, where they might resolve some problems like inflation or late time acceleration (K-inflation \cite{k-infl} or K-essence \cite{k-ess}). Another relevant issue in cosmology is the formation of (topological or non-topological) defects \cite{Vil} - \cite{Dzhu1} where, again, K field theories allow for a much richer phenomenology  \cite{bab-muk-1} - \cite{fring}. Specifically, the formation of domain walls is described by effectively 1+1 dimensional theories \cite{babichev1} - \cite{bazeia3}, \cite{comp}, with possible applications to the structure formation in the early universe. In this context, the problem of supersymmetric extensions of K field theories emerges naturally. Indeed, if the fundamental theory (e.g., string theory) is supersymmetric, and if some of the supersymmetry is assumed unbroken even for the effective field theory in a regime of not too low energy (e.g., in the very early universe \cite{Rocher}, \cite{Yama}, \cite{baumann1}, \cite{baumann2}), then it is an important question whether the resulting supersymmetric effective field theory can be described, at all, in the context of K field theories. The investigation of this problem has been resumed very recently. Concretely, in \cite{ovrut}, supersymmetric (SUSY) extensions of some 3+1 dimensional K field theories with cosmological relevance (ghost condensates, galileons, DBI inflation) have been investigated, whereas the SUSY extensions of some lower-dimensional theories relevant, e.g., for domain wall formation, have been studied in \cite{bazeia2}, \cite{susy-bS}, \cite{SUSY-K-def}. 

If SUSY extensions of some K field theories can be constructed, and if these SUSY K field theories support topological defect solutions, then the following very important questions arise immediately: are the topological defects BPS solutions? And, if so, are they invariant under part of the SUSY transformations? Further, if the defect solutions can be classified by a topological charge, does this charge reappear in the SUSY algebra as a central extension?   All these interrelated features are well-known to show up in SUSY field theories with standard kinetic terms
\cite{divecchia-ferrara}, \cite{witten-olive}, \cite{edelstein-nunez}, \cite{d'adda-horsley}, \cite{d'adda-divecchia}, and SUSY allows, in fact, to better understand both the existence and the structure of BPS solutions. Analogous results for SUSY K field theories would, therefore, be very important for a better understanding of these theories. It is the purpose of the present paper to investigate this question for a large class of SUSY K field theories in 1+1 dimensions.

Concretely, in \cite{SUSY-K-def} we introduced a class of SUSY K field theories and studied their domain wall solutions, but in that paper we were not able to determine whether these topological defects were of the BPS type. As a consequence, all the related questions listed above could not be adressed, either. In the present paper we shall close these loopholes. In Section II, we briefly review the class of SUSY K field theories we consider and, in a next step, demonstrate the BPS property of all their domain wall solutions. In Section III, then, we demonstrate that the domain wall (kink) solutions are invariant under part of the SUSY transformations, and that they show up in the SUSY algebra as central extensions. We also briefly discuss the same issue for the class of models originally introduced in \cite{bazeia2}. Finally, Section IV contains our conclusions.

\section{The BPS bound}

\subsection{The models}

The present paper continues the investigation of the models introduced in \cite{SUSY-K-def}, therefore we use the same conventions as in that reference, to which we refer for details. The field theories we consider exist in 1+1 dimensional Minkowski space, and we use the metric convention $ds^2 \equiv g_{\mu\nu} dx^\mu dx^\nu = dt^2 - dx^2$. Furhter,  we use the superfield ($\theta^2 = \frac{1}{2}\theta^\alpha \theta_\alpha$)
\begin{equation}
\Phi (x,\theta ) =\phi (x) +\theta^\gamma\psi_\gamma (x) -\theta^2 F(x) ,
\end{equation}
and for the spinor metric to rise and lower spinor indices we use $C_{\alpha \beta} = -C^{\alpha\beta}=(\sigma_2)_{\alpha\beta}$. 
For the gamma matrices we choose a representation where the components of the Majorana spinor are real. Concretely, we choose (the $\sigma_i$ are the Pauli matrices)
\be 
\gamma^0 = \sigma_2 \, , \quad \gamma^1 = i\sigma_3 \, , \quad \gamma^5 = \gamma^0 \gamma^1 = -\sigma_1 .
\ee  
Further, the superderivative is
\begin{equation}
D_\alpha=\partial_\alpha+i\theta^\beta\partial_{\alpha\beta}=\partial_\alpha -i \gamma^\mu{}_\alpha{}^\beta \theta_\beta \partial_\mu 
\end{equation}
and allows to extract the components of an arbitrary superfield via ($D^2 \equiv \frac{1}{2} D^\alpha D_\alpha$):
\be
\label{comp}
\phi(x)=\Phi(x,\theta )|,\quad\,\psi_{\alpha}(x)=D_{\alpha}\Phi(x,\theta )|,\quad\,F(x)=D^2\Phi(x,\theta )|,
\ee
(the vertical line $|$ denotes evaluation at $\theta^\alpha =0$). A Lagrangian always is the $\theta^2$ component of a superfield, so it may be calculated from the corresponding superfield via the projection $D^2 \vert$. 

Attempts to find supersymmetric extensions of field theories with nonstandard kinetic terms typically face the problem that the auxiliary field couples to derivatives or becomes dynamical. Recently, however, we found linear combinations of superfields such that the auxiliary field $F$ still obeys an algebraic field equation and, in the bosonic sector, only couples to the scalar field $\phi$ and not to derivatives \cite{SUSY-K-def}. The construction uses the following superfields as building blocks,
\be \label{building-blocks}
{\cal S}^{(k,n)} =  (\frac{1}{2} D^\alpha \Phi D_\alpha \Phi) (\frac{1}{2} D^\beta D^\alpha \Phi
D_\beta D_\alpha \Phi )^{k-1}(D^2 \Phi D^2 \Phi )^n 
\ee
where $k=1,2,\ldots$ and $n=0,1,2,\ldots$. The right linear combinations are
\be 
{\cal S}^{(k)} \equiv \sum_{n=0}^{k-1} (-1)^n \binom{k}{n} {\cal S}^{(k-n,n)}
\ee
and arbitrary linear combinations of these expressions, each one multiplied by an arbitrary real function $\alpha_k (\Phi)$  of the superfield $\Phi$, are permitted. In addition, we may include a superpotential $-P(\Phi)$. That is to say, we define the superfield
\be 
{\cal S}^{(\alpha ,P)} \equiv \sum_{k=1}^N \alpha_k (\Phi) {\cal S}^{(k)} - P(\Phi)
\ee
(here $\alpha = (\alpha_1 ,\alpha_2 , \ldots , \alpha_N )$ is a multiindex of scalar functions),
then the bosonic sector (i.e., with the fermions set equal to zero, $\psi_\alpha =0$) of the corresponding Lagrangian,
\be
{\cal L}^{(\alpha ,P)}_{\rm b} \equiv \left( -D^2 {\cal S}^{(\alpha ,P)}\vert \right)_{\psi =0}
\ee
(b stands for "bosonic") reads explicitly
\be
{\cal L}_b^{(\alpha ,P)} 
= \sum_{k=1}^N \alpha_k(\phi)  [ (\partial^\mu\phi\partial_\mu\phi)^k + (-1)^{k-1}F^{2k}] - P'(\phi) F
\ee
and, as announced, $F$ only appears algebraically and does not couple to derivatives, see \cite{SUSY-K-def} for details.

 In a next step, we should eliminate $F$ via its algebraic field equation
\be \label{F-eq}
\sum_{k=1}^N (-1)^{k-1} 2k \alpha_k (\phi) F^{2k-1} - P' (\phi )=0
\ee
which, however, for a given $P(\phi)$ is a rather complicated equation for $F$ with several solutions. 
It is, therefore, more natural to assume a given on-shell value $F=F(\phi)$ for $F$ and interpret the above equation as a defining equation for the corresponding superpotential $P$.
Eliminating the resulting $P'(\phi)$ we arrive at the Lagrangian density
\bea \label{L-b-F}
{\cal L}_b^{(\alpha ,F)}
&=& \sum_{k=1}^N \alpha_k(\phi) [ (\partial^\mu\phi\partial_\mu\phi)^k - (-1)^{k-1}(2k-1) F^{2k}] 
\eea
where now $F=F(\phi)$ is a given function of $\phi$ which we may choose freely depending on the system we want to study. The $\alpha_k (\phi)$, too, are functions which we may choose freely, but they should obey certain restrictions in order to guarantee, e.g., positivity of the energy, or the null energy condition (NEC), see \cite{SUSY-K-def} for details. 
Next, we have to briefly discuss the field equations.
For a general Lagrangian ${\cal L}(X,\phi)$ where $X\equiv \frac{1}{2} \partial_\mu \phi \partial^\mu \phi = \frac{1}{2}(\dot{\phi}^2 - \phi '^2)$,
the Euler--Lagrange equation reads
\begin{equation}
\partial_\mu ( {\cal L}_{,X} \partial^\mu \phi ) - {\cal L}_{,\phi} =0, 
\end{equation}
 and the energy momentum tensor is
\begin{equation}
T_{\mu\nu} = {\cal L}_{,X} \partial_\mu \phi \partial_\nu \phi - g_{\mu\nu} {\cal L} .
\end{equation}
For static configurations $\phi = \phi (x)$, $\phi ' \equiv \partial_x \phi$, only two components of the energy momentum tensor are nonzero,
\begin{eqnarray} \label{stat-en-de}
T_{00} &=& {\cal E} = -{\cal L} \\
T_{11}  &=& {\cal P} = {\cal L}_{,X} \phi'^2 + {\cal L}
\end{eqnarray}
where ${\cal E}$ is the energy density and ${\cal P}$ is the pressure. Further, for static configurations the Euler--Lagrange equation 
may be integrated once to give
\begin{equation} \label{zero-press}
-2X{\cal L}_{,X} + {\cal L} = \phi '^2 {\cal L}_{,X} + {\cal L} \equiv {\cal P} =0
\end{equation}
(in general, there may be an arbitrary, nonzero integration constant at the r.h.s. of Eq. (\ref{zero-press}), but the condition that the vacuum has zero energy density sets this constant equal to zero).
For the Lagrangian (\ref{L-b-F}), we, therefore, get the once integrated static field equation
\begin{equation} \label{Lb-first-order-eq}
\sum_{k=1}^N (2k-1) (-1)^{k-1}\alpha_k (\phi)  \left( \phi'^{2k} -  F^{2k} \right) =0.
\end{equation}
In a first step, it is useful to interpret this equation as an algebraic, polynomial equation for $\phi'$ of order $2N$. It obviously has the two solutions (roots)
\be 
\phi' = \pm F (\phi)
\ee 
which are independent of the $\alpha_k (\phi)$, therefore we call them "generic" roots. In addition, in general it will have $2N-2$ further roots
\be
\phi' = \pm R_i (\phi) \; ,\quad i=2, \ldots ,N
\ee
(we set $R_1 =F$), which depend both on $F(\phi)$ and on $\alpha_k (\phi )$. We, therefore, call them "specific" roots.

\subsection{Kink solutions}

In a next step, we now interpret the roots $\phi' = \pm R_i (\phi)$ as first order differential equations and want to understand under which conditions their solutions may be topological solitons (kinks and antikinks). A first condition is that the potential term in the Lagrangian (\ref{L-b-F}),
\be \label{V-a-F}
V^{(\alpha ,F)}= \sum_{k=1}^N \alpha_k(\phi)  (-1)^{k-1}(2k-1) F^{2k}
\ee
must have at least two vacua, i.e., field values $\phi = \phi_{0,l}$ such that $V(\phi_{0,l})=0$, where $l=1 , \ldots ,L$ and $L\ge 2$. Now we will make some simplifying assumptions. The functions $\alpha_k (\phi)$ should have no singularities, i.e., $|\alpha_k (\phi)| <\infty$ for $|\phi |  <\infty$, such that no kinetic term gets artificially enhanced. Further, the standard kinetic term should never vanish, i.e., $\alpha_1 >0 \; \forall \; \phi$. Under these assumptions, the standard kinetic term dominates in the vicinity of the vacua, and the standard asymptotic analysis for kink solutions applies.  A kink (antikink) is a static solution $\phi_k (x)$ which interpolates between two vacua, $\phi_k (\pm\infty) \equiv \phi_\pm \in \{ \phi_{0,l} \}$, where for a kink it holds that $\phi_+ > \phi_-$, whereas for an antikink $\phi_- > \phi_+$. We shall assume in what follows that the signs of all the roots $R_i$ have been chosen such that $\phi' =+R_i$ corresponds to the kink (if this equation has a kink solution, at all), and $\phi' =-R_i$ corresponds to the antikink.

A necessary condition for a root $R_i (\phi)$ to provide a kink solution is that it must have two zeros at two different vacua, i.e., $R_i (\phi_\pm)=0$. This is a nontrivial condition because, generically, roots may have no or one zero, as well, with the only condition that the total number of zeros of all the roots coincides with the number of vacua of the potential, including multiplicities. In other words, both the existence of a sufficient number of vacua and the existence of roots with two zeros requires some finetuning of the functions $F$ and $\alpha_k$. The simplest way to achieve this finetuning is via symmetry considerations. If, for instance, $F$ and all the $\alpha_k$ are symmetric under the reflection $\phi \to -\phi$, then all the roots $R_i$ inherit this symmetry. If, therefore, a root has a zero $\phi_{0,l}$ then it has the second zero $-\phi_{0,l}$, by construction. The only additional finetuning required in this case is that the potential must have at least one vacuum at $\phi \not= 0$.

The generic root $\phi' =F$ will lead to a kink solution if the function $F$ has at least two zeros, which obviously provide the corresponding vacua in the potential, see Eq. (\ref{V-a-F}). We shall call the resulting kink solutions "generic kinks". If we choose, e.g., $F=1-\phi^2$, then all models with this $F$ (i.e., for arbitrary $\alpha_k$) will have the standard $\phi^4$ kink $\phi_k =\tanh (x-x_0)$ (here $x_0$ is an integration constant reflecting translational invariance). Depending on the $\alpha_k$, these models may have further kink solutions, based on some of the specific roots $R_i, \; i=2,\ldots ,N$. If these kinks exist, we shall call them "specific kinks". 

We remark that for different roots which only have one zero each, but for different vacuum values, it is sometimes still possible to construct kink solutions interpolating between the two vacua in the space {\bf C}$^1$ of continuous functions with a continuous first derivative. Indeed, if two different roots $R_i$ and $R_{j}$ with two different zeros have a common range of values $\phi \in [\phi_< , \phi_>]$ between the two vacua, then we may form a kink solution in the space {\bf C} of continuous functions with a discountinuous first derivative by joining the two local solutions at any value in the common range (the joining point $x_0$ in base space is arbitrary due to translational invariance). If, in addition, the equation $R_i (\phi) = R_j (\phi)$ has a solution $\phi_s$ in the common range, then the derivatives of the two local solutions coincide at this point, and we may form a kink solution in the space {\bf C}$^1$ by joining the two local solutions at $\phi_s$. Let us point out that if we require kinks to be solutions of the corresponding variational problem, then solutions in the space {\bf C}$^1$ are perfectly valid. They lead to well-defined energy densities and, therefore, provide well-defined critical points of the corresponding energy functional. For more details and some explicit examples, we refer to \cite{SUSY-K-def}.

\subsection{Kink energies and BPS bounds}

In a next step, we want to study the energies of kinks.
The energy density for the Lagrangian (\ref{L-b-F}) is 
\be
{\cal E}_b^{(\alpha ,F)} = \sum_{k=1}^N \alpha_k (\phi) \left( (\dot \phi^2 - \phi '^2)^{k-1} ((2k-1)\dot\phi^2 + \phi '^2) + (-1)^{k-1} (2k-1) F^{2k} \right)  
\ee
and, for static configurations,
\be \label{stat-en}
  {\cal E} = \sum_{k=1}^N (-1)^{k-1} \alpha_k (\phi) \left (\phi '^{2k} + (2k-1) F^{2k}\right) .
\ee
With the help of Eq. (\ref{Lb-first-order-eq}), for kink solutions this may be expressed like
\be \label{w-def}
  {\cal E} = \sum_{k=1}^N (-1)^{k-1} 2k\alpha_k (\phi) \phi '^{2k}  = \phi' \sum_{k=1}^N (-1)^{k-1} 2k\alpha_k (\phi) \phi '^{2k-1} \equiv
\phi' w(\phi ,\phi')
\ee
where the last expression is especially useful for the calculation of the corresponding energies. Indeed, for the energy calculation we should now replace $\phi'$ in $w(\phi ,\phi')$ by the root $R_i$ which corresponds to the kink solution, and interpret the resulting function of $\phi$ as the $\phi$ derivative of another function. That is to say, we define an integrating function $W_i (\phi)$ for each root $R_i$ via
\be \label{W-def}
W_{i,\phi} \equiv w(\phi ,R_i(\phi)) = \sum_{k=1}^N (-1)^{k-1} 2k\alpha_k (\phi) R_i^{2k-1} ,
\ee
then the kink energy is
\be
E=\int dx \phi ' W_{i,\phi} =\int d\phi W_{i,\phi} =W_i (\phi_+) - W_i (\phi_-).
\ee
For the calculation of the kink energy we, therefore, do not have to know the kink solution. We just need the root and the two vacuum values $\phi_\pm$ of the kink. For the {\bf C}$^1$ kinks described above which are constructed by joining local solutions for two different roots $R_i$ and $R_j$, we need the two corresponding integrating functions and the joining point $\phi_s$. The energy then results in
\be
E=W_j(\phi_+) - W_j (\phi_s) + W_i (\phi_s) - W_i (\phi_-).
\ee 
Until now, the energy considerations have been for arbitrary roots, but now we shall see that the generic root $R_1 \equiv F$ apparently plays a particular role. Firstly, the integrating function of the generic root is just the superpotential, $W_1 =P$. Indeed, we find
\be
W_{1,\phi}=  \sum_{k=1}^N (-1)^{k-1} 2k\alpha_k (\phi) F^{2k-1} \equiv P' (\phi)
\ee
see Eq. (\ref{F-eq}). Secondly, if the generic root has a kink solution, then this solution is, in fact, a BPS solution and saturates a BPS bound, as we want to demonstrate now. In general, an energy density has a BPS bound if it may be written off-shell (i.e. without using the static Euler-Lagrange equation) as
\be \label{BPSform}
{\cal E}= (PSD)(\phi ,\phi') + t(x)
\ee
where $(PSD)$ is a positive semi-definite function of $\phi$ and $\phi'$, and $t(x)$ is a topological density, i.e., a total derivative whose integral only depends on the boundary values $\phi_\pm$. Further, a soliton solution (a kink $\phi_k$) is of the BPS type, i.e., saturates the BPS bound if the positive semi-definite function is zero when evaluated for the kink, $(PSD)(\phi_k ,\phi_k')=0$. In our case, the possible topological terms are the expressions $\phi' W_{i,\phi}$ for the different roots. In any case, a possible topological term must be linear in $\phi'$ in order to be a total derivative (we emphasize, again, that the BPS form (\ref{BPSform}) must be valid off-shell, i.e., it is not legitimate to replace $\phi'$ by a root $R_i$ or vice versa). Let us now demonstrate that the energy density may be expressed in BPS form (\ref{BPSform}) for the generic topological term $t=\phi' W_{1,\phi} \equiv \phi' P_{,\phi}$, and that the corresponding positive semi-definite function is zero precisely for the generic kink, i.e., for $\phi' =F$. Indeed, we find for the difference ${\cal E}-t$ for the generic topological term
\bea \label{BPS-firstline}
{\cal E} - \phi' P_{,\phi} &=& \sum_{k=1}^N (-1)^{k-1} \alpha_k (\phi) \left (\phi '^{2k} + (2k-1) F^{2k} - 2k \phi' F^{2k-1} \right) \\
&=& (\phi'-F)^2 S(\phi' ,F) \equiv (\phi' -F)^2 \sum_{k=1}^N (-1)^{k-1} \alpha_k (\phi) H_k (\phi ',F) \label{BPS-secondline}
\eea
where
\be
H_k (\phi ',F) \equiv \sum_{i=1}^{2k-1} i \phi'^{2k-1-i}F^{i-1}.
\ee
Before proving this algebraic identity, we want to make some comments. The above result implies a genuine BPS soliton provided that the positive semi-definite function is zero only iff $\phi$ obeys the corresponding generic kink equation $\phi' =F$. This implies that $S(\phi' ,F)$ must be strictly positive for any nontrivial field configuration (for the trivial vacuum $\phi'=0$ and $F=0$ it holds that $S(0,0)=0$), i.e., $S(a,b)>0$ unless $a=0$ and $b=0$. This inequality, indeed, holds for each individual term $H_k(a,b)$, i.e., $H_k(a,b)>0$ unless $a=0$ and $b=0$ (the proof requires two complete inductions, therefore we relegate it to appendix A). The inequality $S(a,b)>0$ for the complete function $S$, therefore, implies some restrictions on the functions $\alpha_k (\phi)$ (one possible choice is that the $\alpha_k$ are zero for even $k$ and positive semi-definite for odd $k$, but there are less restrictive choices). This is similar to the conditions of positivity of the energy density, or the NEC, which, too,  imply some restrictions on the $\alpha_k$, (again, $\alpha_k$ zero for even $k$ and positive semi-definite for odd $k$ is a possible choice), and we shall assume in the sequel that the $\alpha_k$ obey these restrictions (i.e., the restrictions resulting from the condition $S>0$, and either positivity of the energy density or the NEC; these restrictions are probably related, but we shall not investigate this problem further and assume the two restrictions independently). Now let us prove the algebraic identity between Eq. (\ref{BPS-firstline}) and Eq. (\ref{BPS-secondline}). This follows from the following identities (we set $\phi' =a$, $F=b$)
\bea
&& a^{2k } + (2k-1) b^{2k} -2kab^{2k-1}   \\
&=& (a-b) \left(a^{2k-1} + a^{2k-2}b + a^{2k-3} b^2 + \ldots + ab^{2k-2} - (2k-1) b^{2k-1}\right)  \nonumber\\
&=& (a-b)^2 \left(a^{2k-2} + 2 a^{2k-3} b + 3 a^{2k-4}b^2 + \ldots + (2k-1) b^{2k-2}\right) \nonumber \\
&\equiv & (a-b)^2 H_k (a,b) 
\eea
where the equality of adjacent lines may be checked easily.

So we found, indeed,  that generic kinks (if they exist) saturate a BPS bound, whereas up to now we could not make a comparable statement about additional "specific" kinks. This special role played by the generic kink solution is not surprising from the point of view of the supersymmetric extension, because only the generic kink obeys the simple equation $\phi' =F$, and only the generic kink has a topological charge which may be expressed in terms of the superpotential. On the other hand, the special character of the generic kink {\em is} surprising from the point of view of the purely bosonic theory
\be
{\cal L}_b
= \sum_{k=1}^N \alpha_k(\phi)  (\partial^\mu\phi\partial_\mu\phi)^k -V(\phi)
\ee
(with given $\alpha_k$ and a given potential $V$), whose once-integrated static field equation just leads to the $2N$ roots
\be
\phi' = \pm R_i (\phi) \; , \quad i=1,\ldots ,N
\ee
without distinguishing them in terms of an auxiliary field or a superpotential. The resolution of the puzzle may be understood if we express the once-integrated static field equation both in terms of the potential and in terms of the on-shell auxiliary field,
\begin{equation} \label{Lb-first-order-eq2}
\sum_{k=1}^N (2k-1) (-1)^{k-1}\alpha_k (\phi)  \left( \phi'^{2k} -  F^{2k} \right) = \sum_{k=1}^N (2k-1) (-1)^{k-1}\alpha_k (\phi)   \phi'^{2k} 
-V=0.
\end{equation}
Up to now we assumed a given $F(\phi)$ which lead to the two generic roots $\phi' =F$ and the remaining, specific roots. But now we may interpret this equation in a different way. We may treat only $V$ and the $\alpha_k$ as given and try to find all the solutions for $F$ of the equation
\be
\sum_{k=1}^N (2k-1) (-1)^{k-1}\alpha_k (\phi)  F^{2k}  = V. 
\ee
Obviously, the solutions are just the roots $F_i=R_i (\phi)$ (see Eq. (\ref{Lb-first-order-eq})), and the corresponding first order equations now just read $\phi' =\pm F_i$. We remark that different on-shell choices $F_i$ for the auxiliary field $F$ lead to different superpotentials and, therefore, to different supersymmetric extensions. As a result, the resolution of the puzzle is that one given bosonic theory allows for $N$ different supersymmetric extensions such that each kink solution is the generic solution of its corresponding supersymmetric extension. As a consequence, the energy density allows for BPS bounds for all kink solutions. The existence of several BPS bounds for one and the same energy density may seem surprising, but the different bounds exist, of course, in different topological sectors (i.e., for different boundary values), so there is no contradiction. Finally, all topological charges (i.e., all BPS energies) are now given in terms of the corresponding superpotentials. Indeed, we calculate (see Eqs. (\ref{F-eq}), (\ref{w-def}) and (\ref{W-def}))
\be
W_{i,\phi}(\phi) = w(\phi , R_i(\phi)) = w(\phi ,F_i) = P'(F_i(\phi)) \equiv P'_i (\phi) .
\ee 
We remark that from a practical point of view it is still useful to choose a specific on-shell $F(\phi)$, because in this way we may choose simple functions with simple kink solutions. For generic $\alpha_k$ and a generic $V$, on the other hand, the resulting roots will usually be quite complicated.

\section{SUSY algebra and central extensions}

From now on, we will, again, restrict to a fixed supersymmetric extension, i.e., to fixed, given $\alpha_k$, a fixed, given on-shell auxiliary field $F(\phi)$ and the corresponding superpotential given by Eq. (\ref{F-eq}). The SUSY transformations of the fields read
\be
\delta \phi =  \epsilon^\alpha \psi_\alpha \; , \quad \delta \psi_\alpha = -i(\gamma^\mu)_\alpha{}^\beta \epsilon_\beta \partial_\mu \phi - \epsilon_\alpha F \; , \quad
\delta F = i  \epsilon^\alpha (\gamma^\mu)_\alpha{}^\beta \pa_\mu \psi_\beta 
\ee
(where $\epsilon_\alpha = (\epsilon_1 ,\epsilon_2 )$ are the Grassmann-valued SUSY transformation parameters, and $\epsilon^\alpha = (i\epsilon_2 ,-i\epsilon_1)$), or more explicitly
\bea
\delta \phi &=& i\left( \epsilon_2 \psi_1 -  \epsilon_1 \psi_2\right)  \nonumber \\
\delta F &=& i \left( \epsilon_2 (\psi_1' - \dot{\psi}_2) - \epsilon_1 (\dot{\psi}_1 - \psi_2') \right) \nonumber \\
\delta \psi_1 &=& \epsilon_1 (\phi ' -F) - \epsilon_2 \dot{\phi} \nonumber \\
\delta \psi_2 &=& \epsilon_1 \dot{\phi} - \epsilon_2 (\phi ' +F) .
\eea
Obviously, for a generic kink solution $(\dot \phi =0, \phi ' =F, \psi_\alpha =0)$ the SUSY transformation restricted to $\epsilon_2 =0$ is zero, whereas for a generic antikink the restriction $\epsilon_1=0$ gives zero.

On the other hand, the SUSY transformations of the fields are generated by the SUSY generators $Q=\epsilon^\alpha Q_\alpha$ via the commutators $\delta \phi = [iQ,\phi ]$, etc., where $Q$ should be determined from the Noether current of the SUSY transformations, and the commutators are evaluated with the help of the canonical (anti-)commutation relations of the fields. The supercharges $Q_\alpha$ are known to obey the algebra
\be 
\{ Q_\alpha , Q^\beta \} = 2 \Pi_\nu (\gamma^\nu )_\alpha{}^\beta +2i Z (\gamma^5 )_\alpha{}^\beta
\ee
or, explicitly, 
\bea
Q_1^2 &=& \Pi_0 +Z \nonumber \\
Q_2^2 &=& \Pi_0 - Z  \nonumber \\
\{ Q_1 , Q_2\} &=& 2\Pi_1
\eea
where the curly bracket is the anti-commutator, $\Pi_\nu = (\Pi_0 ,\Pi_1) $ are the energy and momentum operators, and $Z$ is a possible central extension which the SUSY algebra may contain. An explicit calculation of the operators which appear in the SUSY algebra requires the knowledge of the Noether current and the canonical momenta and, therefore, of the complete SUSY Lagrangian, including the fermionic terms, which, in general, is quite complicated. If we only want to determine the central charge, however, it is enough to evaluate the SUSY algebra for a specific field configuration, because the central charge is essentially a number (it commutes with all operators) and, therefore, must take the same value for all field configurations within a given topological sector. We now evaluate the SUSY algebra for a generic kink solution and make the reasonable assumption that not only the restricted SUSY transformation (i.e., the action of the corresponding SUSY charge on the fields) for a generic kink is zero, but that the corresponding SUSY charge itself is zero when evaluated for the generic kink. As we know the energy of the kink, this allows then to determine the central charge. Concretely, for the kink the corresponding charge is $Q_2$, and we get
\be
Q_2^2 = 0 = E_k -Z = P(\phi_+) - P(\phi_-) -Z \quad \Rightarrow \quad Z=P(\phi_+) - P(\phi_-) ,
\ee
where $P$ is the superpotential, and $\phi_\pm$ are the asymtopic values of the kink. For the antikink, $Q_1$ is zero, and we find $Z= P(\phi_-) - P(\phi_+)$. We remark that for positive semi-definite energy densities the resulting restrictions on the functions $\alpha_k$ imply that the central extension $Z$ is always positive, because $P'\ge 0$ for the kink, and $P' \le 0$ for the antikink, as follows from the energy density (\ref{stat-en}) and the defining equation for $P'$, Eq. (\ref{F-eq}). 
We, therefore, found exactly the same result for the central extension as in the case of the SUSY extension of a standard scalar field theory with a quadratic kinetic term for the boson field.

\subsection{Central extensions for the models of Bazeia, Menezes and Petrov}

Here we want to demonstrate that the same central extensions of the SUSY algebra in terms of the superpotential may be found for another class of 
supersymmetric K field theories, originally introduced by Bazeia, Menezes and Petrov (BMP) \cite{bazeia2}. They are based on the superfield
\be 
{\cal S}_{\rm BMP} =  f(\pa_\mu \Phi \pa^\mu \Phi )\frac{1}{2} D_\alpha \Phi D^\alpha \Phi 
\ee
and lead to the bosonic Lagrangian
\be \label{Baz}
{\cal L}_{\rm BMP} 
= f(\pa_\mu \phi \pa^\mu \phi )(F^2 + \pa_\mu \phi \pa^\mu \phi ).
\ee
Here, the Lagrangian produces a coupling of the auxiliary field $F$ with the kinetic term $\pa_\mu \phi \pa^\mu \phi$, but, on the other hand, the auxiliary field only appears quadratically, implying a linear (algebraic) field equation for $F$. 
The same bosonic Lagrangians may, in fact,  be constructed from the building blocks (\ref{building-blocks}) of Section 2 by taking a different linear combination
(the fermionic parts of the corresponding Lagrangians will in general not coincide)
\be
{\cal S}^{(k)}_{\rm BMP} \equiv  \sum_{n=0}^{k-1} (-1)^n \binom{k-1}{n} {\cal S}^{(k-n,n)}
\ee
leading to the bosonic Lagrangians
\be
 {\cal L}^{(k)}_{\rm BMP} = (F^2 + \pa_\mu \phi \pa^\mu \phi ) (\pa_\mu \phi \pa^\mu \phi )^{k-1} .
\ee
We may easily recover the Lagrangian (\ref{Baz}) by taking linear combinations of these,
\be \label{BMP-new}
 {\cal L}_{\rm BMP} = \sum_{k=1}^\infty \beta_k {\cal L}^{(k)}_{\rm BMP} = (F^2 + \pa_\mu \phi \pa^\mu \phi ) \sum_k \beta_k (\pa_\mu \phi \pa^\mu \phi )^{k-1} \equiv (F^2 + \pa_\mu \phi \pa^\mu \phi ) f(\pa_\mu\phi \pa^\mu \phi) .
\ee
Adding a superpotential, the resulting bosonic Lagrangians are
\be
{\cal L}_{\rm BMP}^{(P)} = f(\pa_\mu \phi \pa^\mu \phi ) (F^2 + \pa_\mu \phi \pa^\mu \phi ) - P' (\phi )F,
\ee
or, after eliminating the auxiliary field $F$ using its algebraic field equation
\be \label{BMP-F-eq}
F= \frac{P'}{2f},
\ee
\be
{\cal L}_{\rm BMP}^{(P)} = f \cdot (\frac{P'^2}{4f^2} + \pa_\mu \phi \pa^\mu \phi ) -
\frac{P'^2 }{2f} = f \cdot \pa_\mu \phi \pa^\mu \phi  - \frac{P'^2}{4f}.
\ee
The energy functional for static configurations may be written in a BPS form. Indeed,
\be
{ E}_{\rm BMP}^{(P)} = \int dx \left( \phi '^2 f + \frac{P'^2}{4f}\right) = \int dx \left( \frac{1}{4f}(2\phi ' f \mp P')^2 \pm \phi ' P' \right)
\ee
and for a solution to the first order (or BPS) equation
\be \label{BMP-BPS-eq}
2\phi' (x) f(-\phi'^2) =P'
\ee
 (we take the plus sign for a kink) the resulting energy is  
\be
{ E}_{\rm BMP}^{(P)} = \int_{-\infty}^\infty dx \phi ' P' = \int_{\phi (-\infty)}^{\phi (\infty)} d\phi P' =P(\phi_+) - P(\phi_-).
\ee
Finally, from Eq. (\ref{BMP-F-eq}) for $F$ and the BPS equation (\ref{BMP-BPS-eq}) it follows that the equation $\phi' =F$ still holds for a kink solution and, therefore, the restricted SUSY transformation with only $\epsilon_1$ nonzero is, again, zero when evaluated for the kink. We conclude that the central charge in the SUSY algebra is, again, given by the topological term 
\be
Z=| P(\phi_+) - P(\phi_-) |
\ee
for this class of models.

\section{Conclusions}
In this paper we carried further the investigation of a class of SUSY K field theories originally introduced in \cite{SUSY-K-def}. Concretely, we demonstrated that all the domain wall solutions which exist for this class of field theories are, in fact, BPS solutions. Further, these BPS solutions are invariant under part of the SUSY transformations. We also found strong indications (based on a very reasonable assumption) that the topological charges carried by the domain wall solutions show up in the SUSY algebra as central extensions. That is to say, the situation we found is exactly equivalent to the case of standard SUSY theories with BPS solitons, despite the much more complicated structure of the SUSY K field theories investigated here. Let us emphasize, again, that from an effective field theory point of view, K field theories are as valid as field theories with a standard kinetic term, and there exists no reason not to consider them. Even one and the same topological defect with some given, well-known physical properties may result either from a theory with a canonical kinetic term, or from a certain related class of K field theories (so-called noncanonical twins of the standard, canonical theory), \cite{trodden}, \cite{twin}. K field theories should, therefore, be considered on a par with standard field theories in all situations where they cannot be excluded a priori. This implies that also the study of their possible SUSY extensions is a valid and relevant subject. Structural investigations of the type employed in the present paper are, then, important steps towards a better understanding of these supersymmetric generalized field theories with nonstandard kinetic terms.

\section*{Acknowledgement}
The authors acknowledge financial support from the Ministry of Education, Culture and Sports, Spain (grant FPA2008-01177), the Xunta de Galicia (grant INCITE09.296.035PR and Conselleria de Educacion), the Spanish Consolider-Ingenio 2010 Programme CPAN (CSD2007-00042), and FEDER. Further, AW was supported by polish NCN grant 2011/01/B/ST2/00464.

\section*{Appendix A}
We want to prove that 
\be
a^{2k-2} + 2 a^{2k-3} b + \ldots + (2k-1) b^{2k-2} > 0 \quad \forall \quad k
\ee
unless $a=0$ and $b=0$. For $a=0$, $b\not= 0$, and for $a\not= 0$, $b=0$ this is obvious, so we may restrict to the case $a\not= 0$ and $b\not= 0$. In this case, we may divide by $b^{2k-2}$, so that we have to prove ($x\equiv a/b$)
\be
f_k (x) \equiv x^{2k-2} + 2 x^{2k-3} + \ldots + (2k-1) >0
\ee
which we do by complete induction. Obviously, the statement is true for $k=1$: $f_1 (x) = x^2 + 2x +3=(x+1)^2 +2>0$. Now we assume that it holds for $f_k$ and calculate $f_{k+1}$. We get
\be
f_{k+1}(x) = x^{2k} + 2 (x^{2k-1} + x^{2k-2} + \ldots + 1) + f_k (x) \equiv g_k (x) + f_k (x)
\ee
and the statement is true if $g_k(x) \ge 0 \; \forall \; k$. This, again, we prove by induction. Obviously, it is true for $k=1$: $g_1 (x) = x^2 + 2x +2 \ge 0$. For $g_{k+1}$ we calculate 
\be
g_{k+1} (x) = x^{2k} (x+1)^2 + g_k (x)
\ee
and it is obviously true that $g_k(x) \ge 0 \; \Rightarrow \; g_{k+1}(x) \ge 0$ and, therefore,  $f_k(x) > 0 \; \Rightarrow \; f_{k+1}(x) > 0$, which is what we wanted to prove.

\end{document}